\documentclass[%
 aip,
 amsmath,amssymb,
 preprint,%
]{revtex4-1}

\usepackage{graphicx}% Include figure files
\usepackage{dcolumn}% Align table columns on decimal point
\usepackage{bm}% bold math
\usepackage[mathlines]{lineno}% Enable numbering of text and display math
\usepackage[utf8]{inputenc}
\usepackage[T1]{fontenc}
\usepackage{mathptmx}
\usepackage{etoolbox}

\makeatletter
\def\@email#1#2{%
 \endgroup
 \patchcmd{\titleblock@produce}
  {\frontmatter@RRAPformat}
  {\frontmatter@RRAPformat{\produce@RRAP{*#1\href{mailto:#2}{#2}}}\frontmatter@RRAPformat}
  {}{}
}%
\makeatother
\begin{document}

%\preprint{AIP/123-QED}

%\title[A Vertical GaN $\alpha$-Particle Detector]{Development and Performance Study of Vertical GaN $\alpha$-Particle Detector With High Energy Resolution}
\title{Development and Performance Study of Vertical GaN $\alpha$-Particle Detector with High Energy Resolution}

% Force line breaks with \\
\author{Minjie Ye}
 \affiliation{ Fujian Science \& Technology Innovation Laboratory for Optoelectronic Information of China, Fuzhou 350108, China}%
 \affiliation{Department of Engineering Physics, Tsinghua University, Beijing 100084, China}

\author{Yuzi Yang}%
\email{yangyz18@tsinghua.org.cn}
\affiliation{School of Nuclear Science and Technology \& MOE Frontiers Science Center for Rare Isotopes, Lanzhou University, Lanzhou 730000, China}
\affiliation{Department of Engineering Physics, Tsinghua University, Beijing 100084, China}

\author{Jiangtao Wei}
\affiliation{School of Integrated Circuits, Tsinghua University, Beijing 100084, China}
\author{Weilong Qin}
\affiliation{School of Integrated Circuits, Tsinghua University, Beijing 100084, China}

\author{Hao Hong}
\affiliation{School of Integrated Circuits, Tsinghua University, Beijing 100084, China}%
\affiliation{Department of Microelectronics, Delft University of Technology, 2628 CD Delft, The Netherlands}
\author{Dong Han}
\author{Jianping Ni}
\affiliation{Department of Engineering Physics, Tsinghua University, Beijing 100084, China}
\author{Zhiyi Liu}
\affiliation{School of Nuclear Science and Technology \& MOE Frontiers Science Center for Rare Isotopes, Lanzhou University, Lanzhou 730000, China}
\author{Po-Chung Huang}
 \affiliation{ Fujian Science \& Technology Innovation Laboratory for Optoelectronic Information of China, Fuzhou 350108, China}
\author{Cheng-Chang Yu}
 \affiliation{ Fujian Science \& Technology Innovation Laboratory for Optoelectronic Information of China, Fuzhou 350108, China}
\author{Chao-Yi Fang}
 \affiliation{ Fujian Science \& Technology Innovation Laboratory for Optoelectronic Information of China, Fuzhou 350108, China}
 \author{Entsai Lin}
 \affiliation{ Fujian Science \& Technology Innovation Laboratory for Optoelectronic Information of China, Fuzhou 350108, China}
\author{Zewen Liu}
\affiliation{School of Integrated Circuits, Tsinghua University, Beijing 100084, China}%
\email{liuzw@tsinghua.edu.cn}
\author{Shaomin Chen}
\affiliation{Department of Engineering Physics, Tsinghua University, Beijing 100084, China}
\email{chenshaomin@tsinghua.edu.cn}

\date{\today}

\begin{abstract}
High-energy-resolution GaN $\alpha$-particle detectors have significant potential for space radiation, nuclear instrumentation, and harsh-environment applications. 
However, existing GaN $\alpha$-particle detectors still face several key challenges, including reducing the dead-layer thickness, suppressing leakage current under high reverse bias, improving energy resolution, and clarifying the physical mechanism underlying the low-energy tail phenomenon. 
This study presents a vertical homoepitaxial GaN $\alpha$-particle detector integrating a 20-nm ultrathin dead layer and a guard-ring structure. 
The detector exhibits an ultralow leakage current of 2.195 nA at -200 V and an intrinsic energy resolution of 2.69\% with a charge collection efficiency (CCE) of 95.9\% at -260 V. 
More importantly, this work demonstrates for the first time through Geant4 simulations that depletion‑width nonuniformity is the dominant source of partial energy leakage, leading to an extended low‑energy tail in the energy spectrum.
We establish a depletion-width nonuniformity model and observe good agreement between simulation and experiment. 
This finding provides practical guidance for the design and optimization of high-performance GaN-based radiation detectors.
\end{abstract}

\maketitle
%\tableofcontents

\section{\label{sec:introduction}Introduction}
Gallium nitride (GaN), characterized by its strong Ga–N bond and wide bandgap of 3.4 eV~\cite{Jia2025,Li2026}, has emerged as a promising semiconductor material for radiation detection under extreme environments~\cite{Yasar2025,Lee2012}. Owing to its excellent radiation hardness~\cite{Pei2025,Zhu2018}, high breakdown electric field~\cite{Lian2022}, and superior thermal stability~\cite{Huang2025,Vaitkus2003,Wang2012}, GaN enables the development of radiation detectors capable of operating at high temperatures~\cite{Liu2026}, under high radiation flux~\cite{Zhou2023}, and in harsh environmental conditions~\cite{Kassem2025}. These attributes render GaN-based detectors particularly suitable for $\alpha$-particle monitoring in applications~\cite{Larkin2024}, such as nuclear reactor instrumentation and deep-space exploration, where high energy resolution and high energy spectrum fidelity are critical~\cite{Wang2015,Hou2021}.

To date, GaN $\alpha$-particle detectors can be broadly classified into heteroepitaxial and homoepitaxial devices. Heteroepitaxial GaN detectors grown on foreign substrates typically suffer from high crystal defect densities~\cite{Yao2025,Zhou2025}, with the best reported energy resolution of about 4\%~\cite{Geng2021}. 
In contrast, homoepitaxial GaN detectors benefit from superior crystal quality~\cite{Kum2019}, achieving the best reported energy resolution of 1.8\% and a CCE of 75.7\% at -100 V~\cite{Sandupatla2019}. Nevertheless, these devices still face critical challenges, including elevated leakage current under high reverse bias and considerable energy loss due to thick surface dead layers, which constrain energy resolution and energy spectrum accuracy~\cite{Sandupatla2019}.

Despite continuous progress in device fabrication and material quality, a pronounced low-energy tail remains a frequently observed feature of measured $\alpha$-particle energy spectra from GaN detectors~\cite{Hou2021,Sellin2004,Yang2021,Xu2017}. 
This low-energy tail broadens the spectral peak and degrades the detector's energy resolution~\cite{Pomme2015}, thereby limiting its energy-spectral performance. Previous studies have attributed the low-energy tail to mechanisms such as partial self-absorption of $\alpha$ particles in the source substrate~\cite{Shaikina2026}, air absorption~\cite{Zhang2024}, crystal defects~\cite{Nino2026}, and local difference in carrier collection~\cite{Bai2025}. 
However, these explanations were proposed for other radiation detector systems, whereas the physical mechanism underlying the low-energy tail in GaN $\alpha$-particle detectors remains unclear. Consequently, a quantitative physical model capable of reproducing the experimentally observed spectral asymmetry has not yet been established for GaN detectors.

In this work, a high-performance vertical homoepitaxial GaN $\alpha$-particle detector was designed and fabricated. The detector features a 20-$\mu$m-thick epitaxial layer, a 20-nm ultrathin dead layer, and a guard-ring structure, achieving an intrinsic energy resolution of 2.69\% and a CCE of 95.9\% at -260 V. Furthermore, by correlating capacitance-derived depletion characteristics with Geant4 simulations, we established a depletion-width nonuniformity model to clarify the physical mechanism underlying the low-energy tail phenomenon.

\section{\label{sec:device}Device Fabrication}
\begin{figure*}
\includegraphics{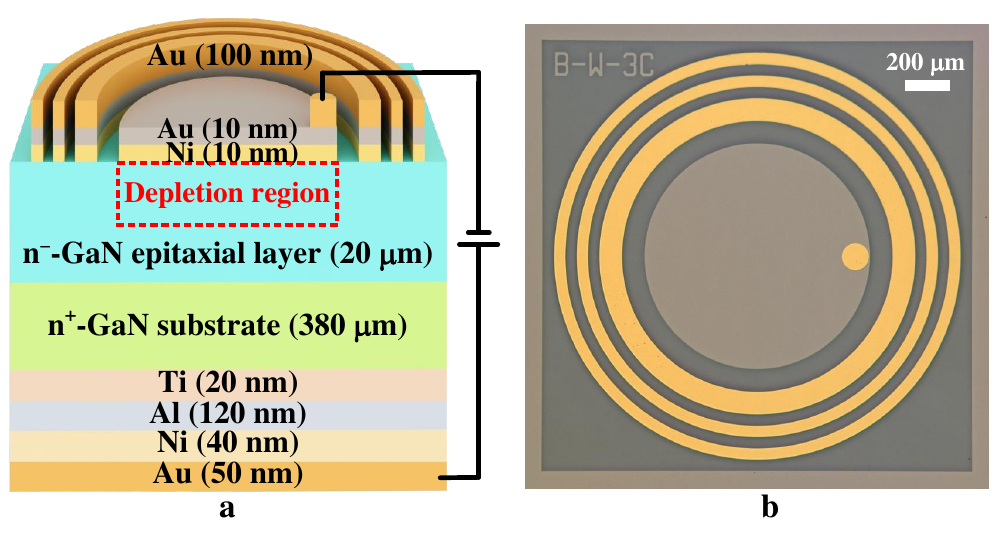}
\caption{\label{fig:GaNDetector}Device structure of the vertical GaN detector. (a): Schematic cross-section of the GaN detector. (b): Optical micrograph of the fabricated device with a 1-mm-diameter Schottky electrode.}
\end{figure*}

FIG.~\ref{fig:GaNDetector}(a) shows the schematic structure of the GaN detector. 
The main fabrication steps are as follows. Firstly, a 20-$\mu$m-thick n$^-$-GaN layer was grown on a 380-$\mu$m-thick n$^+$-GaN substrate with a resistivity of 0.014 $\Omega$·cm by the metal-organic chemical vapor deposition (MOCVD). 
Next, a Ti/Al/Ni/Au (20/120/40/50 nm) metal stack was deposited on the surface of n$^+$-GaN substrate and then annealed in N$_{2}$ ambient at 775 $^{\circ}$C for 45 s to form the Ohmic contact. 
Subsequently, a Ni/Au (10/10 nm) metal layer was deposited on the n$^-$-GaN to define the Schottky contact and the guard ring. Finally, a 120-$\mu$m-diameter, 100-nm-thick Au pad was sputtered onto the Schottky contact region, serving as the wire bonding pad. 
Meanwhile, a 100-nm-thick Au layer was also deposited on the guard-ring structure. When a bias voltage is applied between the Schottky and Ohmic contacts, a depletion region forms in the n$^{-}$-GaN beneath the Schottky contact, defining the sensitive volume in which the charge generated by $\alpha$-particle energy deposition is effectively collected.

The micrograph of the device chip is shown in FIG.~\ref{fig:GaNDetector}(b), and has dimensions of 2 mm $\times$ 2 mm. The Schottky contact electrode, with a diameter of 1 mm and a thickness of 20 nm, forms the detector’s dead layer; reducing its thickness helps minimize the insensitive region. The Au-deposited guard ring effectively suppresses edge leakage current and improves the device’s electrical stability.

\section{\label{sec:Performance}Results and Discussion}
\subsection{\label{Electrica}Electrical measurements}
\subsubsection{Reverse I–V characteristic}
\begin{figure}
\includegraphics{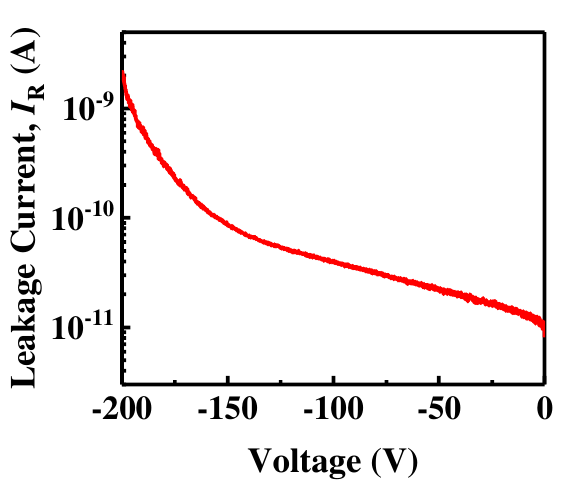}
\caption{\label{fig:IV} Reverse current–voltage characteristics of the GaN detector. Reverse I–V curves measured under dark conditions.}
\end{figure}

The reverse current–voltage (I–V) characteristic of the fabricated vertical GaN $\alpha$-particle detector was measured at 25 $^{\circ}$C using a Keysight B1500A semiconductor parameter analyzer, as shown in FIG.~\ref{fig:IV}. The device exhibits a low leakage current (IR) of 2.195 nA at -200 V.

Such low leakage current is primarily attributed to the high crystalline quality of the homoepitaxial GaN layer and the suppression of edge leakage by the guard-ring structure. Reduced leakage current is essential for minimizing electronic noise and enabling reliable event-by-event pulse-height analysis in $\alpha$-particle energy spectrum measurements.

\subsubsection{Reverse C–V characteristic}
To further characterize the detector’s electrical properties, the reverse capacitance-voltage (C-V) measurement was conducted at a frequency of 1 MHz. For a Schottky diode under reverse bias, the effective net donor concentration (N$_D$) can be extracted from the slope of the linear region of the 1/C$^{2}$-V characteristic according to

\begin{eqnarray}
N_{D}=\frac{2}{q\varepsilon_{0}\varepsilon_{r}S^{2}K_{C}},
\label{eq:ND}
\\
K_{C}=\left| \frac{d(1/C^{2})}{dV} \right|,
\label{eq:IVCV}
\end{eqnarray}
where $S$ denotes the effective area of the Schottky contact. $K_C$ is the slope of the linear fit to the 1/C$^{2}$-V curve, $q$ is the elementary charge, and $\varepsilon_{0}$ and $\varepsilon_{r}$ are the vacuum permittivity and relative dielectric constant of GaN, respectively.
\begin{figure*}
\includegraphics[width=\columnwidth]{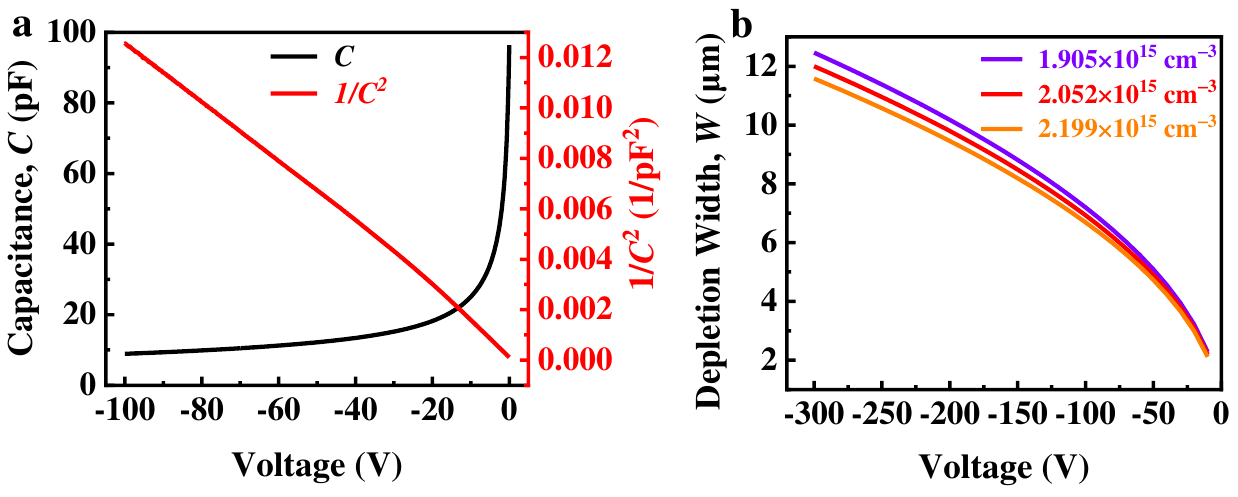}
\caption{\label{fig:CV}Capacitance–voltage analysis of the GaN detector. (a): Measured C–V and 1/C$^2$–V characteristics. (b): Calculated depletion width as a function of reverse bias for representative donor concentrations.}
\end{figure*}

As shown in FIG.~\ref{fig:CV}(a), according to Eq.~\ref{eq:ND}, the effective net donor concentration (N$_D$) was determined to be (2.052 $\pm$ 0.147) $\times$ 10$^{15}$ cm$^{-3}$ from the linear fitting of the 1/C$^{2}$-V curve, where the uncertainty corresponds to the fitting uncertainty derived from the linear regression. This parameter is critical for assessingthe depletion-width distribution and its stability under varying reverse bias, thereby influencing the detector's charge-collection characteristics. 

\subsubsection{Depletion-width evaluation and nonuniformity estimation}
Based on the extracted donor concentration, the depletion width ($W$) of the vertical GaN detector can be evaluated using the one-dimensional depletion approximation:
\begin{equation}
W=\frac{\varepsilon_{0}\varepsilon_{r}S}{C}=\sqrt{\frac{2\varepsilon_{0}\varepsilon_{r}(V_{bi}-V)}{qN_{D}}}
\label{eq:DepW}
\end{equation}
where $V_{bi}$ denotes the built-in potential.

Considering the possible lateral nonuniformity of doping within the area covered by the Schottky contact with a diameter of 1 mm, the effective net donor concentration (N$_D$) was assumed to vary within the range of (1.905 $\sim$ 2.199) × 10$^{15}$ cm$^{-3}$. Three representative values, namely 1.905 $\times$ 10$^{15}$ cm$^{-3}$, 2.052 $\times$ 10$^{15}$ cm$^{-3}$, and 2.199 $\times$ 1015 cm$^{-3}$, were selected to evaluate the sensitivity of the depletion behavior due to variations in doping.

According to Eq.~\ref{eq:DepW}, the calculated depletion width ($W$) increases monotonically when changing reverse bias from -10 V to -300 V for three N$_D$ values, as shown in FIG.~\ref{fig:CV}(b). At a reverse bias of -300 V, the corresponding depletion widths are 12.46 $\mu$m, 11.99 $\mu$m, and 11.59 $\mu$m, respectively. The resulting maximum–minimum depletion width difference induced by the estimated lateral doping variation is therefore $\Delta W$ = 0.87 $\mu$m.

Furthermore, assuming that this depletion-width variation occurs approximately linearly along the lateral direction defined by the Schottky electrode diameter ($D$ = 1 mm), we can estimate the equivalent geometric tilt angle of the depletion boundary as:
\begin{equation}
\theta=\arctan\left( \frac{\Delta W}{D}\right).
\label{eq:theta}
\end{equation}
It yields a value of approximately 0.05$^{\circ}$. This tilt angle provides a physically reasonable representation of lateral depletion-width nonuniformity and serves as a key input parameter for the subsequent particle-transport simulations.

\subsection{$\alpha$-particle energy spectra measurements}
\subsubsection{Experimental setup and measurement conditions}
\begin{figure*}
\includegraphics[width=\columnwidth]{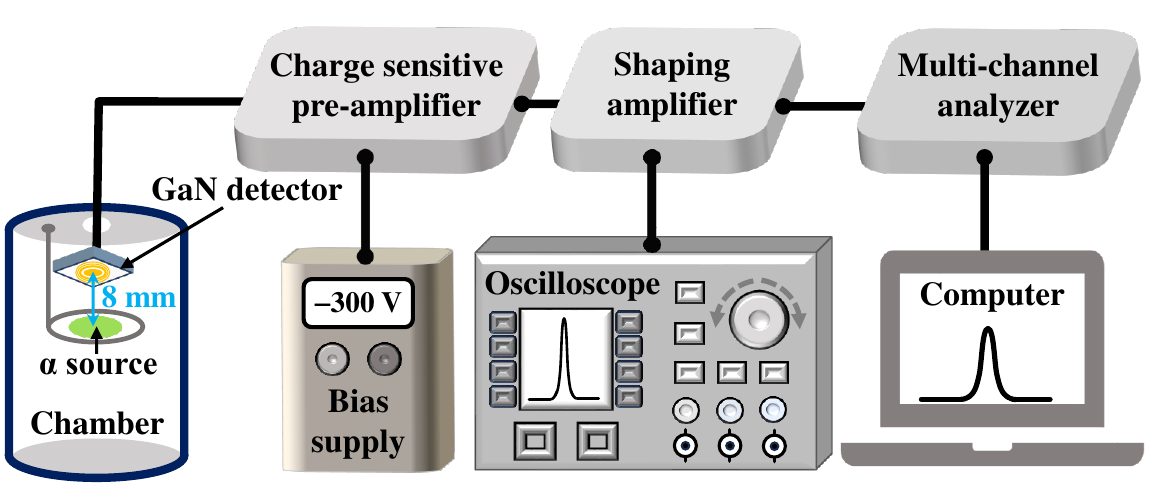}
\caption{\label{fig:alphaMeas}Experimental setup for $\alpha$-particle energy-spectrum measurements. Schematic of the ORTEC-based pulse-processing system used in this study.}
\end{figure*}

To evaluate the energy response characteristics of the fabricated GaN detector to $\alpha$-particles, energy spectra measurements were carried out at 25 $^{\circ}$C using a standard ORTEC modular pulse processing system~\cite{Zhu2018,Sellin2004}, as schematically illustrated in FIG.\ref{fig:alphaMeas}. The measurement system consisted of an $\alpha$ source, a charge-sensitive preamplifier, a bias supply, a shaping amplifier, an oscilloscope, a multi-channel analyzer, and the associated data-processing software.

A $^{241}$Am radioactive source was employed, which predominantly emits $\alpha$-particles with energies of 5.486 MeV and 5.443 MeV, accounting for approximately 98\% of the total emission intensity~\cite{BRADLEY2025,ELLIS1971}. The active area of the source has a diameter of 1 mm, and an $\sim$8 mm air gap was maintained between the source and the GaN detector. 

\subsubsection{Measured energy spectra under different reverse biases}
\begin{figure*}
\includegraphics[width=\columnwidth]{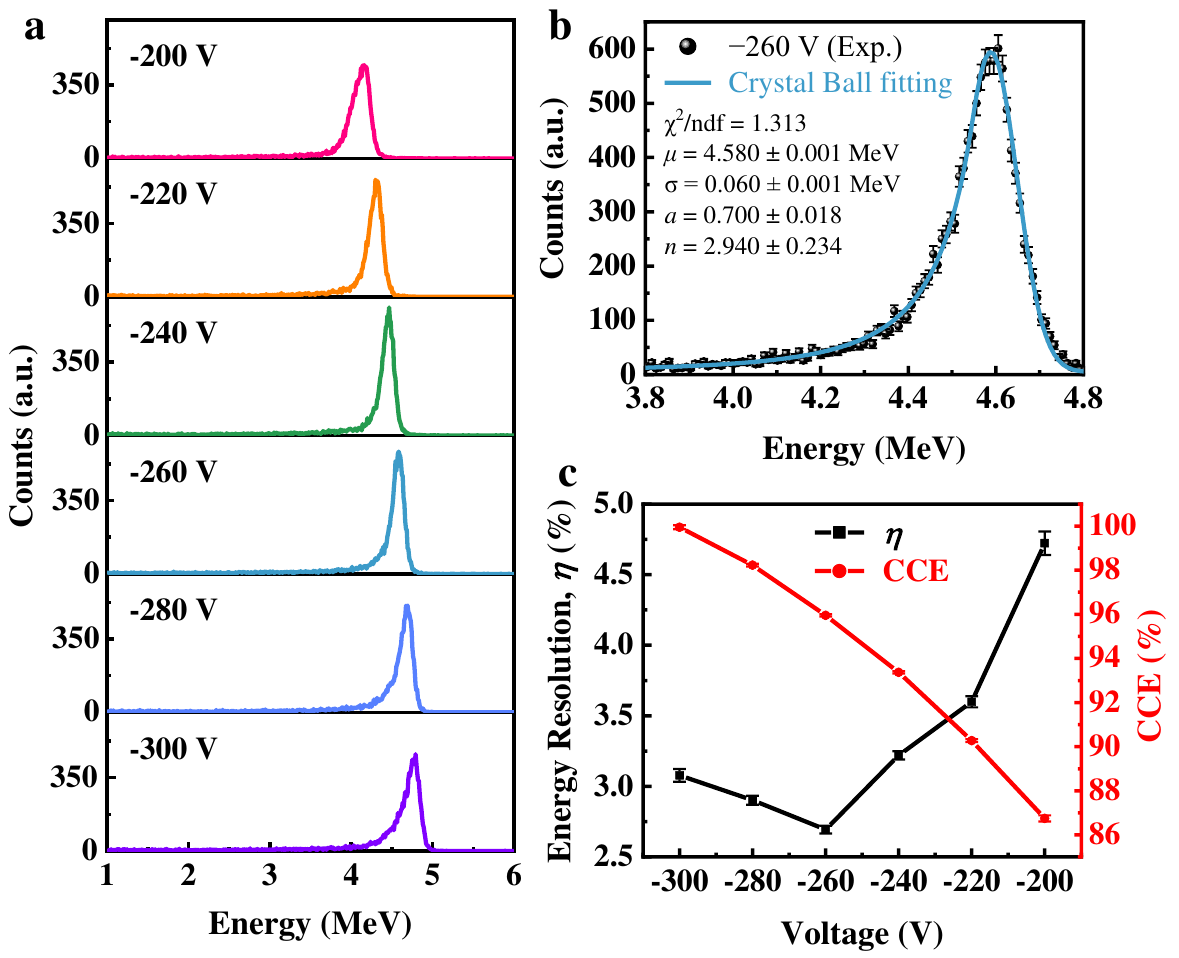}
\caption{\label{fig:MeasResult}Measured energy spectra and bias-dependent performance of the GaN $\alpha$-particle detector. (a): Measured $\alpha$ spectra from $^{241}$Am under various reverse biases. (b): Crystal Ball function fitting to the experimentally measured spectrum at -260 V. (c): Variations of the energy resolution and the CCE as functions of reverse bias.}
\end{figure*}

The energy spectra of the GaN detector to the $\alpha$ source were measured at various bias voltages, and the results are shown in FIG.~\ref{fig:MeasResult}(a). 
As in other experiments~\cite{Hou2021,Xu2017}, distinct low-energy tails are present in all the energy spectra.  
As the absolute value of the reverse bias increases, the peak centroid gradually shifts toward higher energy channels.

This behavior arises from the expansion of the depletion region with increasing reverse bias, thereby extending the effective charge-collection length within the n$^-$-GaN epitaxial layer. As a result, a larger fraction of the $\alpha$-particle energy is deposited and collected within the depleted region, causing the peak centroid to shift toward higher energies.

\subsubsection{Crystal Ball function fitting and low-energy tail description}
To quantitatively characterize the asymmetric spectral shape, the Crystal Ball function~\cite{gaiser_1982,DayaBay:2016ziq} was employed to fit the measured energy spectra. This function combines a Gaussian core with a power-law tail and is widely used to describe radiation-induced energy loss effects. The fitting function is expressed as:
\begin{equation}
f(x;\mu,\sigma,a,n)=N\times \left\{
\begin{aligned}
  &\exp\left(-\frac{(x-\mu)^2}{2\sigma^{2}} \right), & \text{for} \frac{x-\mu}{\sigma}>-a \\
  &A\cdot \left( B-\frac{x-\mu}{\sigma} \right)^{-n}, &\text{for} \frac{x-\mu}{\sigma}\leq -a
\end{aligned}
\right.
\label{eq:csf}
\end{equation}
where
\begin{equation}
A=\left(\frac{n}{a} \right)^{n}\exp\left(-\frac{a^{2}}{2}\right),
\label{eq:csf1}
\end{equation}
\begin{equation}
B=\frac{n}{a}-a,
\label{eq:csf2}
\end{equation}
and $N$ is the normalization factor. The parameters $\mu$ and $\sigma$ represent the peak centroid and standard deviation of the Gaussian component, respectively, while $n$ and $a$ ($a>0$) determine the shape and relative strength of the low-energy tail.

The transition point between the Gaussian core and the power-law tail is located at $\mu$ - a $\times$ $\sigma$. 
A larger value of a shifts the transition point further away from the peak centroid, corresponding to a weaker low-energy tail contribution, whereas a smaller a indicates a more pronounced low-energy tail. For example, the energy spectrum measured at -260 V is shown in FIG.~\ref{fig:MeasResult}(b), demonstrating excellent agreement between the experimental spectrum and the fitting function.

\subsubsection{Energy resolution and charge collection efficiency}
The intrinsic energy resolution ($\eta$) of the detector was evaluated from the full width at half maximum (FWHM) of the Gaussian component, defined as:
\begin{equation}
\eta=2.355\times\sqrt{\left(\frac{\sigma}{\mu}\right)^{2}-\left(\frac{\sigma_{air}}{\mu_{air}}\right)^{2}}
\label{eq:resolution}
\end{equation}
where the 2.355 is the factor that converts standard deviation ($\sigma$) to FWHM, the values of $\sigma$ and $\mu$ from the fitting result from Eq.~\ref{eq:csf}, and the $\frac{\sigma_{air}}{\mu_{air}}$ reflects the energy broadening of the $\alpha$-particle source caused by air. Based on our simulation and calculation, the average residual energy of the $\alpha$-particles reaching the detector surface was determined to be 4.774 $\pm$ 0.031 MeV, accounting for energy loss in air.

FIG.~\ref{fig:MeasResult}(c) shows the intrinsic energy resolution and the CCE as a function of reverse bias. At a reverse bias of -260 V, the detector achieves a high intrinsic energy resolution of 2.69\% together with a high CCE of 95.9\%.

\begin{figure}
\includegraphics[width=0.75\columnwidth]{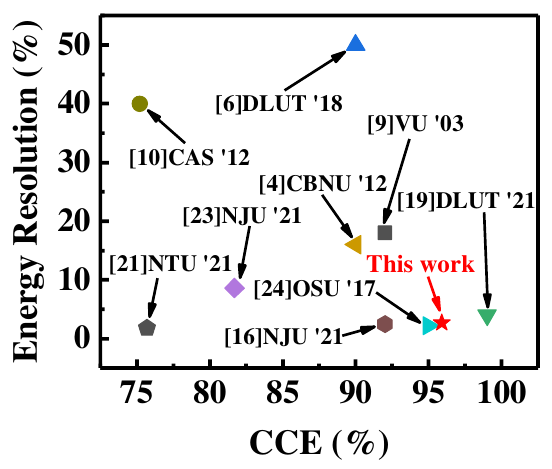}
\caption{\label{fig:Comparison}Benchmarking of GaN-based $\alpha$-particle detectors. Comparison of reported GaN-based $\alpha$-particle detectors in terms of energy resolution and charge collection efficiency.}
\end{figure}

To benchmark device performance, the results are compared with previously reported GaN-based $\alpha$-particle detectors, as summarized in FIG.~\ref{fig:Comparison}. The present device demonstrates internationally advanced performance in both energy resolution and charge collection efficiency.

\subsection{Low-energy tail study}
\subsubsection{Physical origin and modeling approach}

Theoretically, the depletion width of vertical GaN $\alpha$-particle detectors is primarily determined by donor concentration and applied reverse bias. 
However, fluctuations introduced during epitaxial growth and device processing can lead to nonuniformity in the depletion region.
Such nonuniformity is expected to play a critical role in shaping the detector's energy response characteristics.

\begin{figure}
\includegraphics{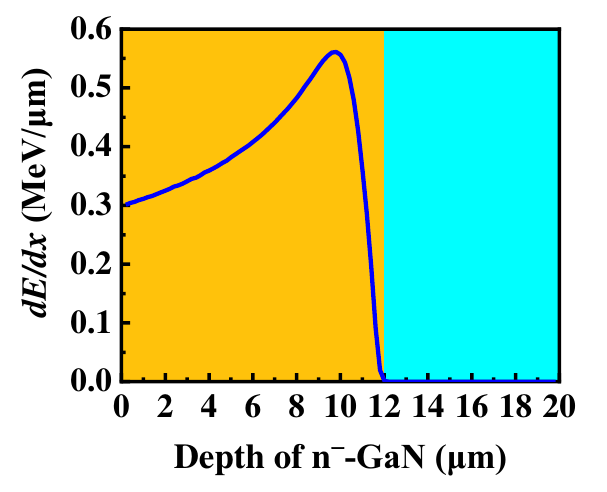}
\caption{\label{fig:GaNDep}Simulated energy-deposition profile of $\alpha$ particles in GaN. $\alpha$ particles emitted from $^{241}$Am in the n$^-$-GaN epitaxial layer after traversing 8 mm of air.}
\end{figure}

FIG.~\ref{fig:GaNDep} presents the simulated energy deposition profile of an $\alpha$ particle emitted from the $^{241}$Am source after traversing 8 mm of air and entering the n$^-$-GaN epitaxial layer, obtained using SRIM~\cite{Ziegler2010SRIM}. 
The simulation reveals a projected range of approximately 12 $\mu$m in n$^-$-GaN, beyond which the $\alpha$-particles are completely stopped and deposit their full residual energy.

In regions where the local depletion width is smaller than the penetration depth of incident $\alpha$-particles, a portion of the deposited energy cannot be fully collected within the depleted volume, resulting in partial energy leakage. This effect manifests as an asymmetric low-energy tail in the measured energy spectrum. To quantitatively investigate this mechanism, particle-transport simulations based on the Geant4 toolkit~\cite{allison_2006,allison_2016} were performed.

\subsubsection{Geant4 simulation model}

Geant4 is a widely adopted Monte Carlo simulation framework for modeling particle transport and energy deposition in matter~\cite{Borexino:2017mly,Lin:2022htc}. In this work, a three-dimensional detector geometry model consistent with the experimental device structure was established, as schematically illustrated in FIG. ~\ref{fig:GaNDetector}(a). The simulation incorporates a GaN epitaxial layer, a Schottky contact, a guard-ring structure, and an air region between the $\alpha$-particle source and the detector surface.

\begin{figure*}
\includegraphics[width=\columnwidth]{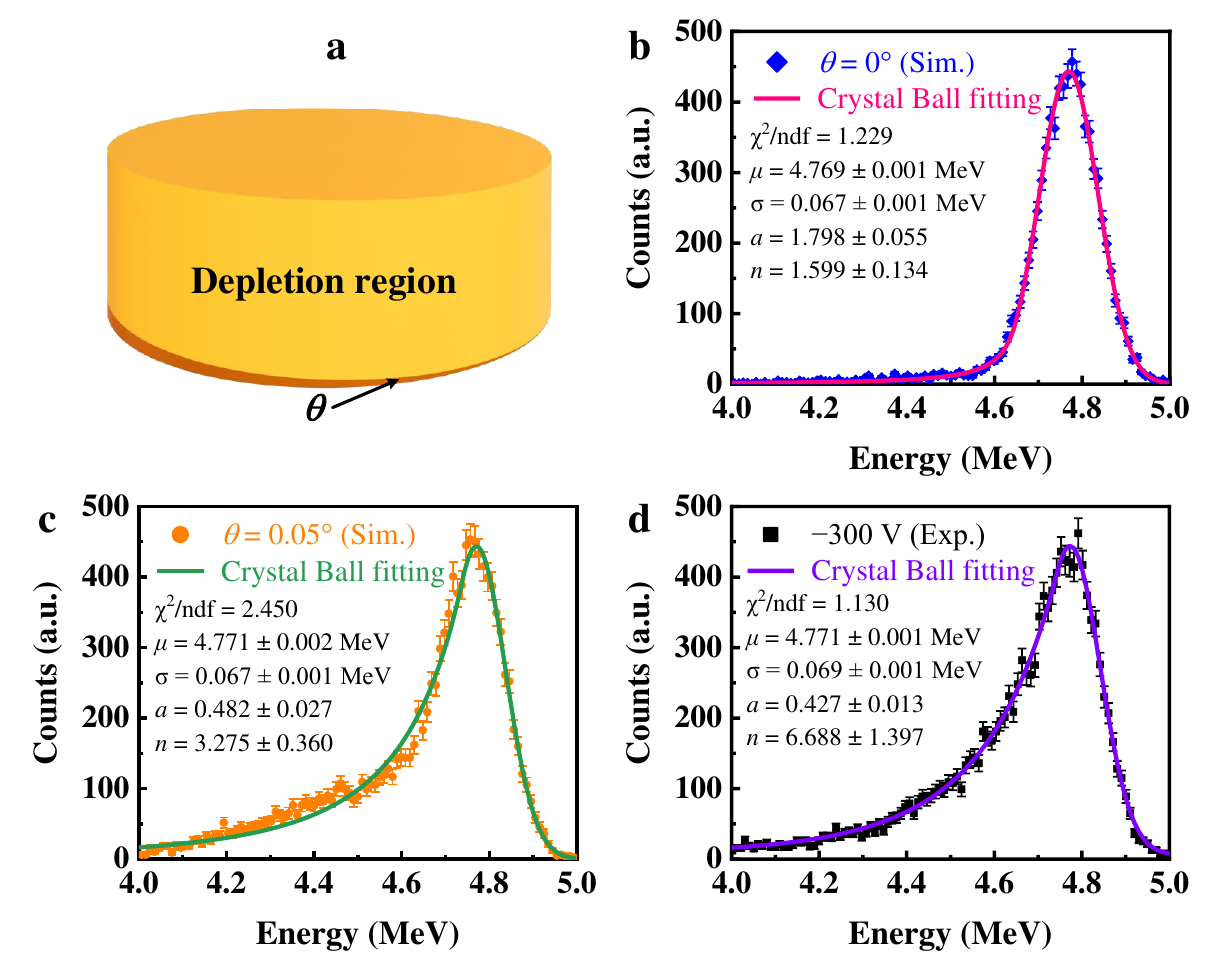}
\caption{\label{fig:Simulation}Depletion-width nonuniformity model and energy spectra. 
(a): In Geant4, a simplified inclined depletion region structure with adjustable tilt angle $\theta$. 
(b): Simulated energy deposition distribution and fitting when $\theta$ = 0$^{\circ}$. 
(c): Simulated energy deposition distribution and fitting when $\theta$ = 0.05$^{\circ}$. d: Measured energy deposition distribution and fitting at -300 V.}
\end{figure*}

To represent depletion-width nonuniformity, an inclined depletion interface was introduced within the n$^-$-GaN layer, as shown in FIG.~\ref{fig:Simulation}(a). 
The tilt angle ($\theta$) of this interface can be adjusted, allowing controlled simulation of the degree of nonuniformity. This simplified representation enables a systematic investigation of energy deposition behavior under varying degrees of nonuniformity.

Throughout the simulations, key parameters were maintained identical to the experimental conditions, including the 1 mm diameter of the $^{241}$Am $\alpha$-particle source, the emission geometry, the 8 mm air gap, and the energy distribution of the emitted $\alpha$-particles.

\subsubsection{Energy deposition behavior and Energy leakage}

When the depletion region is spatially uniform ($\theta$ = 0$^{\circ}$), the maximum depletion width can reach 12.46 $\mu$m under -300 V reverse bias. 
Since the depletion width at all lateral positions exceeds the $\alpha$-particle penetration depth, the energy is fully deposited within the depleted region. 
As shown in FIG.~\ref{fig:Simulation}(b), the resulting energy spectrum exhibits a symmetric Gaussian distribution with negligible low-energy tail contribution.
As shown in FIG.~\ref{fig:Simulation}(b), the resulting energy spectrum presents a relatively symmetrical Gaussian distribution, with the low-energy tail contributing very little.

When depletion-width nonuniformity is introduced ($\theta$ = 0.05$^{\circ}$), the depletion width gradually varies from 11.59 $\mu$m to 12.46 $\mu$m along the lateral direction, consistent with the experimentally estimated depletion-width variation derived from C–V measurements. In regions where the local depletion width is smaller than the $\alpha$-particle range, partial energy deposition occurs outside the depleted volume, leading to incomplete charge collection.

\begin{figure}
\includegraphics{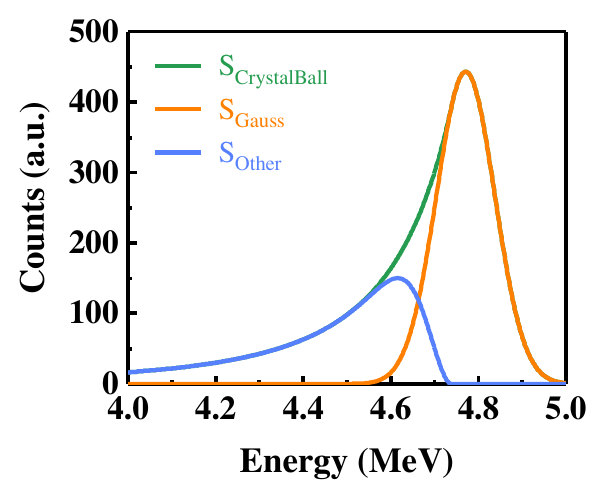}
\caption{\label{fig:EnergyLoss} The decomposition of the Crystal Ball function ($S_\text{CrystalBalls}$). The parameters of the Gaussian component ($S_\text{Gauss}$) are directly taken from the Crystal Ball function itself, while the remaining part ($S_\text{Other}$) is defined as the difference between the Crystal Ball function and the Gaussian component.}
\end{figure}

As shown in FIG.~\ref{fig:Simulation}(c), this localized energy leakage produces a pronounced low-energy tail in the simulated energy spectrum, while the peak centroid remains largely unchanged. 
We attempted to decompose the energy spectrum into two components: the global part $S_{\text{Gauss}}$ and the residual $S_{\text{Other}}$, as indicated by the orange and blue lines in FIG.~\ref{fig:EnergyLoss}. 
The peak, mean ($\mu$), and sigma ($\sigma$) of $S_\text{Gauss}$ are obtained from a fit using the Crystal Ball function ($S_\text{CrystalBall}$) in FIG.~\ref{fig:Simulation}(c). $S_\text{Other}$ is then defined as: 
\begin{equation}
S_\text{Other}=S_\text{CrystalBall}-S_\text{Gauss}
\label{eq:energyloss}
\end{equation}
In this decomposition, $S_{Gauss}$ represents the portion of complete energy deposition, whose distribution is influenced solely by energy resolution, while $S_\text{Other}$ reflects energy loss. Numerical analysis shows that within the 4–5 MeV range, the energy‑loss component ($S_\text{Other}$) accounts for approximately 41.2\%.

Moreover, FIG.~\ref{fig:Simulation}(c) shows that the simulated energy spectrum for ($\theta$ = 0.05$^{\circ}$) exhibits a distinct low-energy tail, which closely matches the experimental result at -300 V in FIG.~\ref{fig:Simulation}(d). 
The two energy spectra are relatively consistent in terms of peak centroid, spectral width, and tail asymmetry, demonstrating that depletion-width nonuniformity is the dominant source of the observed low-energy tail. Together, the simulation and experimental results validate the proposed depletion-width nonuniformity model.

Additional simulations further indicate that other potential mechanisms, including $\alpha$-particle energy loss in the Schottky electrode, absorption by the guard-ring structure, and the finite solid angle of the radioactive source, do not produce a comparable long-tail feature. These effects mainly contribute to minor peak broadening and can therefore be excluded as the primary cause of the spectral asymmetry.

\section{Conclusion}

In summary, a vertical homoepitaxial GaN $\alpha$-particle detector integrating a 20-nm ultrathin dead layer and a guard-ring structure was developed, achieving a high intrinsic energy resolution of 2.69\% and a high charge collection efficiency of 95.9\% at -260 V. 
Beyond the performance improvement, this work clarifies the physical origin of the low-energy tail frequently observed in GaN $\alpha$-particle detectors. By correlating capacitance-derived depletion characteristics with Geant4 simulations, a depletion-width nonuniformity model was established, and, for the first time, this nonuniformity was demonstrated to be the dominant mechanism of partial energy leakage and low-energy tail formation. 
%Quantitative analysis further indicates that improving depletion-width uniformity can reduce energy leakage by approximately 41.2\% in the -300V bias. 
Quantitative analysis further shows that at a bias of –300 V, with a maximum depletion width of 12.46 $\mu$m and a tilt angle of 0.05$^\circ$, the energy leakage caused by depletion-width nonuniformity is approximately 41.2\%.
%Quantitative analysis further indicates that improving depletion‑width uniformity can reduce energy leakage by approximately 41.2\% at a bias of –300 V.
The agreement between simulation and experiment confirms the validity of the proposed physical model. 
These results provide both physical insight and valuable guidance for controlling epitaxial growth and optimizing the structural design of high-performance GaN-based radiation detectors.

\section{Acknowledgements}
This work was supported in part by the National Natural Science Foundation of China (NSFC) under Grant 12127808 and 12505130, the Natural Science Foundation of Fujian Province of China (2026J0011684), and the China Postdoctoral Science Foundation (Certificate Number: 2024M751611).

\nocite{*}
\bibliography{aipsamp}% Produces the bibliography via BibTeX.

\end{document}